\documentclass[12pt]{article}

\usepackage{amsfonts}
\usepackage[english]{babel}
\usepackage{graphicx}
\usepackage{pstricks}

\renewcommand{\thefootnote}{\arabic{footnote}}
\def\ga{\mathrel{\raise.3ex\hbox{$>$\kern-.75em\lower1ex\hbox{$\sim$}}}}
\def\la{\mathrel{\raise.3ex\hbox{$<$\kern-.75em\lower1ex\hbox{$\sim$}}}}
\def\fr#1#2{\frac{#1}{#2}}
\newcommand{\nc}{\newcommand}
\nc{\be}{\begin{eqnarray}}
\nc{\ee}{\end{eqnarray}}

\def\ol{\overline}
\def\beq{\begin{equation}}
\def\eeq{\end{equation}}
\def\beqa{\begin{eqnarray}}
\def\eeqa{\end{eqnarray}}
\def\pref#1{(\ref{#1})}

\def\ca{{\cal A}}

\def\cl{{\cal L}}

\def\sss{\scriptscriptstyle}
\def\ms{m_{\sss S}}
\def\mw{m_{\sss W}}
\def\mz{m_{\sss Z}}
\def\mn{m_{\sss N}}
\def\mp{m_p}
\def\mpl{M_{\rm Pl}}
\def\SM{{\sss SM}}
\def\EW{{\sss EW}}
\def\mh{m_h}
\def\mht{m_{\tilde{h}}}
\def\sw{s_w}
\def\cw{c_w}
\def\ls{\lambda_{\sss S}}
\def\lh{\lambda_h}
\def\l{\lambda}
\def\vrel{v_{\rm rel}}
\def\vew{v_\EW}

\begin{document}

%----------------------------------------------------------------------%
%  numbering equations with section number
%----------------------------------------------------------------------%
\makeatletter
\@addtoreset{equation}{section}
\makeatother
\renewcommand{\theequation}{\thesection.\arabic{equation}}
%----------------------------------------------------------------------%
%  title page
%----------------------------------------------------------------------%
\pagestyle{empty}
%\vspace*{1.0in}
\rightline{TPI-MINN-00/46, UMN-TH-1922-00}
\rightline{McGill-00/31;  IASSNS-HEP-00/83}
\rightline{\tt hep-ph/0011335 }
\vspace{0.5cm}
\begin{center}
\large{\bf The Minimal Model of Nonbaryonic Dark Matter: \\
A Singlet Scalar \\[5mm] }
\large{C.P. Burgess${}^{a,b}$, Maxim Pospelov${}^c$  and
Tonnis ter Veldhuis${}^c$ 
\\[4mm]}
\small{
${}^a$ The Institute for Advanced Study, Princeton, NJ 08540, USA
\\[1mm]
${}^b$ Physics Department, McGill University,\\[-0.3em]
3600 University St.,
Montr\'eal, Qu\'ebec, Canada, H3A 2T8.
\\[1mm]
${}^c$ Department of Physics, University of Minnesota\\
Minneapolis, MN 55455, USA\\[5mm]}
\small{\bf Abstract} \\[5mm]
\end{center}

\begin{center}
\begin{minipage}[h]{14.0cm}
We propose the simplest possible renormalizable extension of the Standard 
Model --- the addition of just one singlet scalar field --- as
a minimalist model for non-baryonic dark matter. Such a model is characterized 
by only three parameters in addition to those already appearing within 
the Standard Model: a dimensionless self-coupling and a mass for the new 
scalar, and a dimensionless coupling, $\l$, to the Higgs 
field. If the singlet is the dark matter, these parameters are related to one 
another by the cosmological abundance constraint, implying that the
coupling of the singlet to the Higgs field is large, 
$\lambda \sim O(0.1 - 1)$. 
Since this parameter also
controls couplings to ordinary matter, we obtain predictions for the 
elastic cross section of the singlet with nuclei. The
resulting scattering rates are close to current limits from both
direct and indirect searches. The existence of
the singlet also has implications for current Higgs searches, as
it gives a large contribution to the invisible Higgs width for much of  
parameter space. 
These scalars can be strongly self-coupled in the cosmologically 
interesting sense recently proposed by Spergel and Steinhardt, but
only for very low masses ($\la 1$ GeV), which is possible only at
the expense of some fine-tuning of parameters.

\end{minipage}
\end{center}
\newpage
%----------------------------------------------------------------------%
%  Resetting of counters
%----------------------------------------------------------------------%
\setcounter{page}{1}
\pagestyle{plain}
\renewcommand{\thefootnote}{\arabic{footnote}}
\setcounter{footnote}{0}
%----------------------------------------------------------------------%
%  Paper begins
%----------------------------------------------------------------------%

\section{Introduction}

It is an amazing fact of our times that even as our understanding of
cosmology progresses by leaps and bounds, we remain almost completely
ignorant about the nature of most of the matter in the universe. 
According to recent fits to cosmological parameters \cite{fit}, dark matter of
some sort makes up close to 30\% of the total energy density. This is
much more than what is inferred from inventories of the luminous matter 
we can see. 
Moreover, the successes of big bang nucleosynthesis suggest that only a
fraction of this matter can be made of ordinary baryons, like
massive compact objects, faint stars, {\it etc.}. Thus, unless
gravity undergoes some drastic changes at distances larger than a few kpc 
(which is quite improbable from several points of view), we must
postulate the existence of enormous amounts of dark matter of
an unknown, non-baryonic origin.

A simple argument shows why current theories of particle physics
are so prolific in suggestions for what the nature of this dark matter might
be \cite{GJK}. The vast majority of the proposed alternatives to the 
Standard Model
involve new particles having masses which are of order $\mw \sim 100$ GeV,
and which couple with electroweak strength to the ordinary matter which
we know and love. If any of these particles is stable enough to
have a lifetime as long as the age of the universe, it makes a natural
candidate for dark matter. It does so because its abundance is naturally
of the right order of magnitude so long as its interaction cross
sections have weak-interaction strength. The abundance comes out right
because it is set by the annihilation rate for particles which are
initially in thermal equilibrium with ordinary matter. Cosmologically
interesting abundances follow pretty much automatically for particles
whose mass is of order $\mw$ and whose annihilation cross sections have
weak (or rather milliweak) interaction strength. (We give this argument in
more detail within the body of the paper.)

Better yet, supersymmetric models, which are perhaps the best motivated
of the many theories which have been proposed, very often have such
long-lived states, due to the natural existence there of a conserved quantum
number, $R$-parity, which keeps the lightest $R$-odd state from decaying. 
These particles cry out for interpretation as dark-matter particles, and 
it is no surprise that these models are by far the most widely explored 
in the literature \cite{susyguys}. 

Best of all, this explanation of the nature of dark matter can be tested
experimentally. This is the direct goal of dedicated dark-matter 
detectors \cite{Cad}, and
an indirect goal of accelerator searches for events with missing energy,
showing that a weakly-interacting particle has escaped the detector. 
If Nature smiles on us we soon may be treated to the discovery of new physics
in both of these kinds of experiments. Indeed, recently the DAMA collaboration 
has announced the detection of a dark matter signal, as indicated by their
seeing an annual modulation of the counting rate in a NaI detector 
\cite{DAMAm}. 
However, the comparably precise data from the Ge detectors of the CDMS 
collaboration \cite{CDMS} do not support these findings. (These two 
experiments need not be in contradiction with each other if
the spin-dependent part of the cross-section is enhanced relative to the
spin-independent part \cite{PV}.) 

Our goal in the present paper is to present a slightly unorthodox view. 
Although very well motivated, supersymmetric models are very complicated
and enjoy an enormous parameter space. This makes them unable 
to definitively predict
what dark-matter detectors must see. Furthermore, unlike the extensive evidence
for the existence of dark matter, the arguments in favour of supersymmetry are
almost exclusively theoretical. In our opinion, with the advent of good-quality
data from dark-matter detectors, it behooves theorists to propose simple models
for the dark matter which are consistent with present evidence, but which make
definite predictions and so are easily falsifiable. These provide benchmarks
against which other models and the data can be compared. 
We believe that it is only by comparing the implications of
such models with one another, and with supersymmetry, that one can hope 
to properly interpret the data. 

The model we study in this paper was first introduced
by Veltman and Yndurain \cite{Velt} in a different context. Its
cosmology was later studied
by Silveira and Zee \cite{Zee}, and (with a complex
scalar) by McDonald \cite{McD}.
It is the absolute minimal modification of the
Standard Model which can explain the dark matter. 
It consists of the addition of
a single spinless species of new particle, $S$, to those 
of the Standard Model, using
only renormalizable interactions. To keep the new particle 
from interacting too strongly
with ordinary matter, it is taken to be completely neutral 
under the Standard Model
gauge group. Besides involving the fewest new states, the model is also just
complicated enough to offer interestingly rich dark matter properties. Unlike
the case if only spin-half or only spin-one singlet particles are added, 
it is possible for a singlet scalar to have {\it both} significant 
renormalizable
self-interactions {\it and} renormalizable interactions with some 
Standard Model fields.

There is also a sense in which the model we propose is generic, 
should the dark
matter consist of a single species of spinless particle. 
To this end, it is useful to ask the question of what a 
generic dark-matter model
should look like. It is clear that the main property which one needs to
ensure is the stability of the new particle, suggesting 
that the fields $S_{i}$,
representing these particles appears in the Lagrangian in even powers, so
that its decay is forbidden. If this field $S_{i}$ is considerably lighter
than the rest of the other exotic undiscovered particles, these may 
be integrated
out, leaving an effective Lagrangian at electroweak scale which has the
generic form
\begin{equation}
{\cal L}={\rm kinetic~term~for}~S_{i}+S_{i}^{*}MS_{i}+\sum
S_{i}^{*}O_{\sss SM}S_{i}+...
\end{equation}
where the kinetic, mass terms and interactions with the SM (via the set of
operators $O_{\sss SM}$) in general would depend on the spin of $S_{i}$. 
The most important couplings at low energies are those of lowest dimension, 
corresponding to the lowest-dimension choices for the operators $O_{\sss SM}$.
Our model also has this form, with only a single singlet scalar $S$. In this
language, our dropping of all nonrenormalizable interactions corresponds
to keeping only those interactions which are consistent with (but do not
require) all other exotic particles to be arbitrarily heavy compared with
the weak scale. We might expect our model to therefore capture the physics
of any more complicated theory whose impact on the dark matter problem
is conveyed purely through the low-energy interactions of a single spinless
particle. 

An additional, more tentative, incentive for formulating more models
stems from recent indications of problems with subgalactic
structure formation within the non-interacting cold-dark-matter scenario 
\cite{problem}. A `generalized' form of cold dark matter may avoid these 
problems if its self-interactions\footnote{
The required self-interactions however also lead to spherical 
halo centers in clusters, which are inconsistent with the
ellipsoidal centers indicated by strong gravitational lensing data 
\cite{Yoshida}.}
can produce scattering cross sections
of order $10^{-24}$ cm${}^2$ in size \cite{selfint}. Within the present
context this proposal would require the masses of dark matter particles 
not to exceed 1 GeV. Within the minimal model described in this paper, 
we find this range of masses may be just barely possible, but requires 
unnatural fine tunings due to the relationship between masses
and couplings imposed by the maintenance of the correct cosmic abundance
of dark-matter scalars.

This paper is organized as follows. In the next section we identify the three
parameters which describe the model, 
and determine the general conditions which lead to 
acceptable masses and to sufficiently 
stable dark matter. In section 3 we calculate the annihilation cross section
of $S$-particles and give the resulting cosmic abundance as a 
function of masses and couplings. 
This calculation is similar to the analysis of Ref. \cite{McD}. 
We perform the numerical analysis for 
the most interesting part of 
the parameter space, with 100 GeV $\le \mh\le$ 200 GeV and 10 GeV $\le \ms \le
$ 100 GeV.
In section 4 we obtain the  cross section for 
elastic scattering 
with ordinary matter and apply the constraints, imposed by direct and indirect
searches. Section 5 computes the cross sections 
for the missing energy events which
are predicted for colliders due to the pair production of $S$ particles. 
It also contains a prediction for the degradation of the Higgs boson signal
at hadronic colliders, when the Higgs boson is allowed to decay into a pair of 
$S$ particles.
Our
conclusions are reserved for section 6.

\section{The Model's Lagrangian}

The lagrangian which describes our model has the following simple
form:
\beq
\label{Ldef}
\cl = \cl_\SM + \frac12 \partial_\mu S \, \partial^\mu S 
- \frac{m_0^2}{2} \; S^2 - \frac{\ls}{4} \, S^4 - \l \, S^2 \, H^\dagger H,
\eeq
where $H$ and $\cl_\SM$ respectively denote the Standard Model 
Higgs doublet and lagrangian,
and $S$ is a real scalar field which does not transform under the
Standard Model gauge group. (Lagrangians similar to this have been 
considered as models for strongly-interacting dark 
matter \cite{BBRT} and as potential complications for Higgs 
searches \cite{invisible}. The same number of free parameters 
appears in the simplest 
Q-ball models \cite{Qball}.) We assume $S$ to be the only new degree of freedom
relevant at the electroweak scale, permitting the neglect of
nonrenormalizable couplings in eq.~\pref{Ldef}, which contains all
possible renormalizable interactions consistent with the field
content and the symmetry $S \to - S$. 

Within this framework the properties of the field $S$ are described
by three parameters. Two of these, $\ls$ and $m_0$ are internal to
the $S$ sector, characterizing the $S$ mass and the strength of its 
self-interactions. Of these, $\ls$ is largely unconstrained and can
be chosen arbitrarily. We need only assume it to be small enough to
permit the perturbative analysis which we present. Couplings to all
Standard Model fields are controlled by the single parameter $\l$.

We now identify what constraints are implied for these couplings by
general considerations like vacuum stability or from the requirement that
the vacuum produce an acceptable symmetry-breaking pattern. These are
most simply identified in unitary gauge, $\sqrt2 \, H^\dagger = (h,0)$
with real $h$, where the scalar potential takes the form:
\beq
\label{Shpot}
V = \frac{m_0^2}{2} \, S^2 + \frac{\l}{2} \; S^2h^2 + \frac{\ls}{4} \, S^4
+ \frac{\lh}{4} \Bigl(h^2 - \vew^2 \Bigr)^2.
\eeq
$\lh$ and $\vew = 246$ GeV are the usual parameters of the Standard Model
Higgs potential.

{\it 1. The Existence of a Vacuum:} 
This potential is bounded from below provided that the quartic
couplings satisfy the following three conditions:
\be
\label{vacstability}
\ls,~\lh &\geq 0& \qquad \hbox{and} \\\nonumber
\qquad \ls \, \lh &\geq& \l^2 \qquad \hbox{for negative $\l$}.
\ee
We shall assume that these relations are satisified and study the minima 
of the scalar potential.

{\it 2. Desirable Symmetry Breaking Pattern:} 
We demand the minimum of $V$ to have the following two properties:
It must spontaneously break the electroweak gauge group, $\langle
h \rangle \ne 0$; and it must not break the symmetry $S \to - S$,
so $\langle S \rangle = 0$. The first of these is an obvious requirement
in order to have acceptable particle masses, while the second
is necessary in order to ensure the longevity of $S$ in a natural way.
($S$ particles must survive the age of the universe in order to play 
their proposed present role as dark matter.) 

The configuration $h \ne 0$ and $S = 0$ is a stationary
point of $V$ if and only if $\vew^2 > 0$, in which case
the extremum occurs at $h_{\rm ext}^2 = \vew^2$. This is a local
minimum if and only if
\be
m_0^2 +\l \, \vew^2 > 0.
\label{localm}
\ee
A second local minimum, with $h_{\rm ext} = 0$ and 
$S_{\rm ext}^2 = - m_0^2/\ls$, 
can also co-exist with the desired minimum if $\l > 0$ and $\l^2 < \lh \ls$. 
This second minimum is present so long as $m_0^2 < 0$ and 
$-\l m_0^2 > \ls \lh \vew^2$. 
Even in this case, the minimum at $S_{\rm ext} = 0$ and 
$h_{\rm ext}^2 = \vew^2$ 
is deeper, and so is the potential's global minimum, provided that
\be 
 0 < - m_0^2 < \vew^2 \sqrt{\lh\ls} .
\ee

Throughout the rest of this paper, the above conditions are assumed
to hold, so that the model is in a phase having potentially acceptable 
phenomenology. It is therefore convenient to shift $h$ by its vacuum
value, $h \to h + \vew$, so that $h$ represents the physical Higgs having
mass $\mh^2 = \lh \, \vew^2$. The $S$-dependent part of the scalar
potential then takes its final form
\beq
\label{Vfinal}
V = \frac12 \, \Bigl(m_0^2 + \l \, \vew^2 \Bigr) S^2 + \frac{\ls}{4} \, S^4 
+ \l \vew \, S^2 \, h + \frac{\l}{2} \, S^2 \, h^2 , 
\eeq
and the $S$ mass is seen to be $\ms^2 = m_0^2 + \l \vew^2$. Our prejudice
in what follows is that this mass lies in the range from a few to a few 
hundred GeV, in which case the resulting dark matter will be cold. 

\section{Constraints from Cosmological Abundance}

We next sharpen the cosmological constraints on the model by demanding
the present abundance of $S$ particles to be close to today's preferred
value of $\Omega_{\sss S}h^2$. This imposes a strong relationship
between the parameters $\l$ and $\ms$, which we now derive. 

We start by assuming that the $S$ particles are in thermal equilibrium
with ordinary matter for temperatures of order $\ms$ and above. This
is ensured so long as the coupling $\l$ is not too small. Just how small
$\l$ must be is determined by the following argument. Thermal equilibrium
requires the thermalization rate, $\Gamma_{\rm th}$, to be larger than
the universal Hubble expansion rate, $H$. The constraint on $\l$ comes from
the demand that this be true throughout the thermal history of the universe,
down to temperatures $T < \ms \sim \mw$. But $\Gamma_{\rm th}$ and $H$ vary 
differently
with time as the universe expands, because they differ in
their temperature dependence. On one hand, in the radiation-dominated epoch
which is of primary interest to us, $H \sim T^2/\mpl$, where $\mpl$ denotes
the Planck mass. 
On the other hand, the thermalization rate 
varies as $\Gamma_{\rm th} \sim \l^2 T$
for $T \gg \mh$, and as $\Gamma_{\rm th} \sim 
\l^2 T^5\mh^{-4}$ for $T \ll \mh$. 
These temperature dependences imply the ratio 
$R_{\rm th} = \Gamma_{\rm th} / H $ is
maximized when $T \sim \mh \sim \mw$, taking the 
maximum value $R_{\rm th}
\sim \l^2 \mpl/\mh$. $S$ particles are therefore guaranteed 
to remain thermalized (or get thermalized) down
to the electroweak epoch, $T \sim \mh \sim \mw$, if this maximum 
ratio is required
to be of order one or larger, implying
\beq
\label{eqcond}
\l \ga \sqrt{\frac{\mw}{\mpl}} \sim 10^{-8}. 
\eeq

Once thermalization is reached (and in the absence of $S$ decays), 
as we shall henceforth assume, the primordial $S$ abundance is determined 
by the $S$ particle mass and its annihilation cross section. 
This cross section depends very strongly on the 
unknown Higgs mass, and on which annihilation channels are 
kinematically open. 
%in a way we now describe. 

An independent, interesting issue is
the fate of a scalar condensate that might
survive from the inflationary epoch. After the Hubble rate drops below
$m_S$, coherent time oscillations of the singlet field begin. These
oscillations can be regarded as the oscillations of a Bose condensate 
of $S$ particles
which is not in thermal equilibrium with other matter.
The fate of the condensate depends on the initial value of the
$S$ field and two possibilities must be distinguished. If the initial value 
of the condensate is sufficiently small so that
the energy density in the oscillations, $\sim m_S^2 \langle S\rangle
^2$,
is smaller than the energy density of radiation, $\sim T^4$, then
the thermalization of this condensate occurs exponentially fast. The
rate is given by $\lambda^2 T$ or $\lambda_S^2 T$, whichever 
is larger. When the
initial value of the condensate is of the order of the electroweak
v.e.v.,
the condensate will therefore completely disappear if 
$\lambda$ or $\lambda_S$ is larger
then $\sqrt {M_W/M_{\rm Pl}}$, just as in the thermalization condition (3.1).
The situation is quite different when the initial value of the $S$ field
is very large ($M_{\rm Pl}$, for example). The $S$ condensate then
dominates the energy density in the Universe and it behaves exactly as the
inflaton
condensate.  The absence of the direct decay of $S$ particles
in this case may prevent the universe from reheating \cite{LKS}
\footnote{We thank Lev Kofman for pointing out this possibility}.
In this paper we assume that $S$ field does not drive inflation, and we
limit ourselves to the first possibility.

%%%%%%%%%%%%%%%%%%%%%%%
\begin{figure}[!t]
\begin{center}
\includegraphics[width=8.0cm]{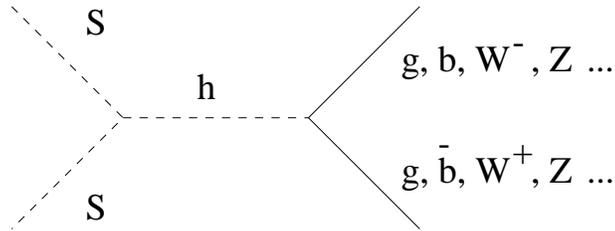}
%{\normalsize \mbox{\epsfxsize=80mm\epsffile{annih.eps}} }
\caption{The Feynman graph relevant to $S$-particle annihilation 
via Higgs exchange. Various annihilation channels are open or 
forbidden, depending on the value of $2\ms$.}
\end{center}
\end{figure}
%%%%%%%%%%%%%%%%%%%%%%%

Since the temperature domain for which annihilation is most 
important is $T_{\rm ann} \sim 0.05 \, \ms$, it is the nonrelativistic
annihilation cross section which is relevant. In our model the
expression for the annihilation rate depends on which phase within 
which it occurs. If it occurs within the Higgs phase, {\it i.e.}
if $T_{\rm ann}$ is low enough so that it occurs after the
electroweak phase transition, the result is given by evaluating the 
tree-level graph of fig.~(1) for $s$-channel annihilation, $SS\to X$,which 
in the nonrelativistic limit gives 
\begin{eqnarray}
\label{approx}
\sigma_{\rm ann} \; v_{rel} &=& 
\frac{8\lambda^{2}\vew^{2}}{(4\ms^{2}-\mh^{2})^{2}+\mh^{2}
\Gamma_{h}^2 } \; F_{\sss X} \\
\hbox{with} \qquad F_{\sss X} &:=& \lim_{\mht \to 2 \ms} 
\left(\frac{\Gamma_{\sss \tilde{h} X}}{\mht}\right) .\nonumber
\end{eqnarray}
Here $\Gamma_h$ is the total Higgs decay rate, and $\Gamma_{\sss \tilde{h} X}$ 
denotes the partial rate for the decay, $\tilde{h} \to X$, for a
virtual Higgs, $\tilde{h}$, whose mass is $\mht = 2 \ms$. Eq. (\ref{approx})
also assumes that $\ms<\mh$, so that direct (contact) annihilation to a pair
of physical Higgses via the $\l S^2h^2$ interaction term is forbidden. 

Of particular interest are the large- and small-$\ms$ limits. The small-$\ms$ 
limit of eq.~\pref{approx} -- $\ms \ll \mw,~\mh$ -- implies the asymptotic 
behaviour: 
\be
\sigma _{\rm ann} v_{rel}
\propto \fr{\lambda^2 \, \ms^2}{\mh^4} \qquad \hbox{if}\; 
\ms \ll \mh,
\label{asympforms1}
\ee
and the coefficient of proportionality depends strongly on the 
accessibility of certain decay channels ($X = \pi\pi$ or 
$\mu^+\mu^-$, and so on).

For large $\ms$ the dominant contributions to the annihilation cross section 
come from $W^+W^-$, $ZZ$ and $hh$ final states. (The latter 
originates from the $\l S^2 h^2$ interaction term, whose contribution must
be summed with (\ref{approx})). Neglecting terms which are $O(\vew^2/\ms^2)$
in the result, we find the large-$\ms$ behavior of the annihilation cross 
section to be   
\be
\label{asympforms}
\sigma _{\rm ann} v_{rel} \approx \fr{\lambda^2}{4\pi\ms^2}, 
\qquad \hbox{if}\; \ms \gg \mh .
\ee
These asymptotic forms are useful in what follows
for understanding what the
cosmological abundance constraint implies for the coupling
$\l$ in the limit where $\ms$ is very large or very small. 
Our results for the annihilation cross section agree with 
the calculation of ref. \cite{McD}.
%%%%
%%%%   
%%%%

%%%%%%%%%%%%%%%%%%%%%%%
%%%%%%%%%%%%%%%%%%%%%%%
\begin{figure}[!t]
\begin{center}
\includegraphics[width=13.0cm]{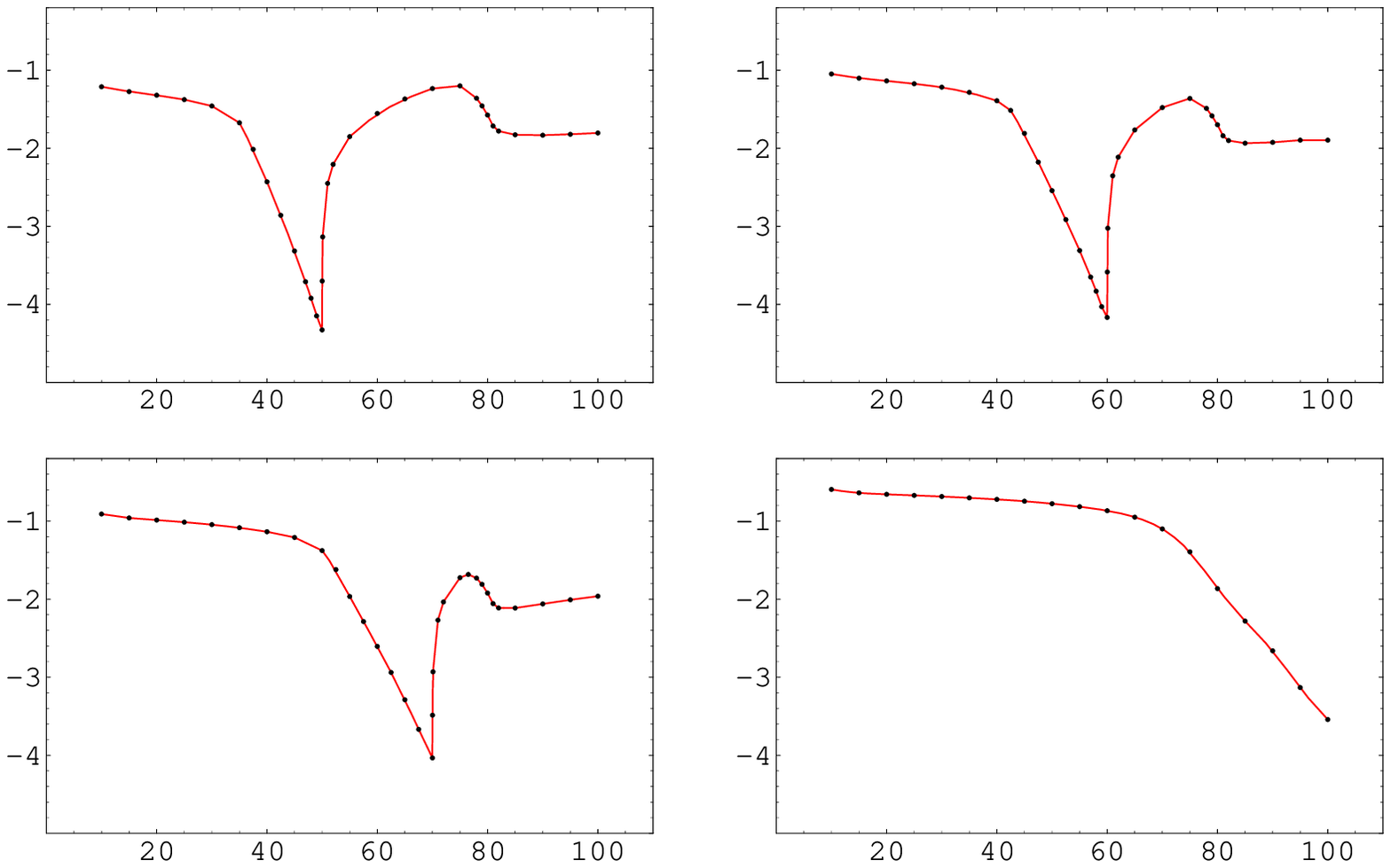}
\put(-199,80){\scriptsize{$m_S \, \left[ {\rm GeV} \right]$}}
\put(-199,0){\scriptsize{$m_S \, \left[ {\rm GeV} \right]$}}
\put(-69,80){\scriptsize{$m_S \, \left[ {\rm GeV} \right]$}}
\put(-69,0){\scriptsize{$m_S \, \left[ {\rm GeV} \right]$}}
\put(-240,97){\scriptsize{$m_h=100\, {\rm GeV}$}}
\put(-240,17){\scriptsize{$m_h=140\, {\rm GeV}$}}
\put(-110,97){\scriptsize{$m_h=120\, {\rm GeV}$}}
\put(-110,17){\scriptsize{$m_h=200\, {\rm GeV}$}}
\put(-258,119){\scriptsize{\rotateleft{$\log\lambda$}}}
\put(-258,39){\scriptsize{\rotateleft{$\log\lambda$}}}
\put(-129,39){\scriptsize{\rotateleft{$\log\lambda$}}}
\put(-129,119){\scriptsize{\rotateleft{$\log\lambda$}}}
\caption{Four samples of the $\log\l$--$\ms$ relationship 
between $\l$ and $\ms$, which gives the correct cosmic
abundance of $S$ scalars. For these plots the Higgs mass is chosen 
to be 100, 120, 140, and 200 GeV. The abundance is chosen to be
$\Omega_sh^2=0.3$}
%
%\psgrid(0,0)(-14,9)(1,-1)
\label{fig1}
\end{center}
\end{figure}

These expressions may be used with standard results for the Standard
Model Higgs decay widths to predict how the primordial $S$-particle
abundance depends on the parameters $\ms$ and $\l$. Standard 
procedures \cite{KT} give the present density of $S$ particles to be
\begin{equation}
\Omega_{s}h^2 = \frac{(1.07\cdot 10^9) \; x_f}{g_*^{1/2}~\mpl ~{\rm GeV}
~\langle \sigma \vrel \rangle}.
\label{abundance}
\end{equation}
Here $g_*$, as usual, counts the degrees of freedom in equilibrium
at annihilation, $x_f = \ms/T_f$ is the inverse freeze-out 
temperature in units of $\ms$, which can be determined by solving 
the equation $x_f \simeq \ln\Bigl[ 0.038 (g_*x_f)^{-1/2} \mpl \, 
\ms \langle \sigma v\rangle \Bigr]$. Finally,
$\langle \cdots \rangle$ denotes the relevant thermal average.

The abundance constraint is obtained by requiring $\Omega_s h^2$ to be
in the cosmologically preferred range, 
which imposes a relation between $\l$ and $\ms$. For our numerical 
results we use $\Omega_s h^2 = 0.3$, which corresponds to a large
value for $\Omega_s = 0.6$, perhaps the largest possible value consistent with
observations. We choose such a large $\Omega_s$ 
in order to be conservative in our later predictions for the signals to 
be expected in dark matter detectors, since larger $\Omega_s$ corresponds
to smaller $\l$. Choosing instead the central value, $\Omega_s h^2 \simeq 0.15$,
obtained from various cosmological fits would give values for $\l^2$ which 
are about twice as large as those which we use in what follows.
Since it happens that $x_f\sim 20$ is approximately
constant for the range of cross-sections and masses expected in 
this model, the abundance condition is equivalent to holding 
$\langle \sigma v\rangle$ constant as $\l$ and $\ms$ are varied
(provided $g_*$ is held constant). 

Fig. (2) plots the relationship between $\l$ and $\ms$
which is predicted in this way by the
requirement that $\Omega_s h^2 \sim 0.3$. For most values of $\ms$ this
curve is well described by the above simple formulae, which give sufficient 
accuracy in most parts of 
the parameter space defined by varying $\ms$, $\l$ and $\mh$. Important
exceptions to this statement apply in kinematically special regions, such as 
the Higgs threshold ($2 \ms \simeq \mh$) and two-particle thresholds in the
final states ($2 \ms \simeq 2 m_b$ or $2 \ms \simeq 2 \mw$, and so on), 
where a more sophisticated treatment is required. In our quantitative
work we follow the numerical procedure outlined in Ref. \cite{GS}
in these special regions.

For $\ms$ sufficiently large or small, the asymptotic expressions
\pref{asympforms1} and \pref{asympforms} show that the abundance constraint
forces $\l$ to become large, eventually becoming too large to believe
perturbative expressions like eq.~\pref{approx}. 
In particular, if annihilation should occur 
before the electroweak transition, then
the asymptotic relation between $\l$ and 
$\ms$ becomes: 
\begin{equation}
\lambda \sim \frac{\ms}{10~{\rm TeV}} ,
\end{equation}
so demanding the perturbative regime ($\l \la 1$) gives the upper
bound $\ms \la 10$ TeV. 

For small $\ms$ we consider the case $\ms=500$ MeV, for which the 
dominant annihilation channels are $\pi^+\pi^-$, $\pi^0\pi^0$ and $\mu^+\mu^-$
\cite{Voloshin}. In this case $\Omega_s h^2 \la 0.3$ is achieved 
if the coupling satisfies the constraint $\l\ga 2$. Smaller 
values of $\l$ for this range of $\ms$ would lead to over-abundant
scalar particles and an overclosure of the universe. 

There are several important points concerning the abundance condition
which bear emphasis:

\begin{enumerate}
\item For all $\ms<{\rm few}$ TeV (and away from poles and 
particle thresholds) 
the abundance constraint requires $\lambda \sim O(0.1 - 1)$. In this sense 
this model of dark matter is ``natural'', in that obtaining the right
primordial abundance does not require any fine tuning or special choice 
of the parameters. (This property is shared by many models for which the
dark matter is a weakly-interacting species of particle whose mass
is of order $\mw$.) 

\item  The coupling $\l$ has to be significantly suppressed (down to
the level of $10^{-4}-10^{-3}$) near the Higgs pole. This is because the Higgs
resonance is rather narrow, and this narrowness considerably enhances 
the $S$ annihilation rate, especially if $2\ms$ is slightly smaller
than $\mh$ \cite{GS}.

\item  Different decay channels dominate the total annihilation 
cross section for different ranges of $\ms$. However, the
range of values of most experimental interest lies between
the $b$ and $W$ thresholds, for which it is
the $b\bar b$ final state that is most important.

\item  Since the abundance constraint is concerned with the strength
of the interactions between $S$ scalars and ordinary matter, it is
largely independent of the strength of the $S$ self-coupling, $\ls$.
This leaves $\ls$ completely free to be adjusted. Unfortunately, although
the $S$ particles therefore can be very strongly interacting, this in
itself does not make them useful to solve the recently-perceived
problems with galaxy formation \cite{problem}. This is because
the solution of these problems requires interaction cross sections
which are of order $10^{-24}$ cm${}^2$, and cross sections this large
require $\ms \la 1$ GeV in addition to large $\ls$ \cite{selfint,BBRT}.
Unfortunately masses this small require fine tuning in this model, due to
the relation $\ms^2 = m_0^2 + \l \vew^2$. Since, as we saw earlier, the 
abundance constraint requires $\l$ to be of order one or larger for small $\ms$,
we require a part-per-million cancellation between  
$\lambda \vew^2$ and $m_0^2$ in order to obtain small values for $\ms^2$. 

It would be interesting to explore whether more natural possibilities
are possible with a less draconianly minimal scalar model. One possibility might
be enhance the cosmic abundance using couplings to SM fermions, while at the same 
time suppressing the coupling to the Higgs to avoid the problem 
of fine-tuning. An alternative approach is to arrange for relatively light, but
strongly self-interacting scalars whose coupling to visible 
matter is suppressed below the level required by the thermalization 
argument which led to eq.~(\ref{eqcond}), and instead arrange for the 
correct cosmic abundance of $S$-particles in another way, perhaps as the
result of an earlier inflationary epoch \cite{FP}. 

\end{enumerate}

\section{Implications for Dark Matter Detectors}

The connection between $\l$ and $\ms$ derived from the abundance constraint
in the previous section is very predictive. In this section we explore its
consequences for current dark matter searches, while the next section 
describes implications for Higgs physics in collider experiments.

The sensitivity of dark matter detectors to $S$ particles is controlled by
their elastic scattering cross section with visible matter, and with nuclei
in particular. This cross section enters in one of two ways: ($i$) it is
the directly relevant quantity for experiments designed to measure the
recoil signal of dark-matter collisions within 
detectors \cite{DAMAm,CDMS,Dama,UKdata,HDM,Zarag}; ($ii$) it 
controls the abundance of dark matter particles which become trapped at
the terrestrial or solar core, and whose presence is detected indirectly
through the flux of energetic neutrinos which is produced by subsequent
$S$-particle annihilation \cite{Kam,Mac,Bak,Amanda,SuperK}.

%%%%%%%%%%%%%%%%%%%%%%%
\begin{figure}[hbtp]
\begin{center}
\includegraphics[width=6.0cm]{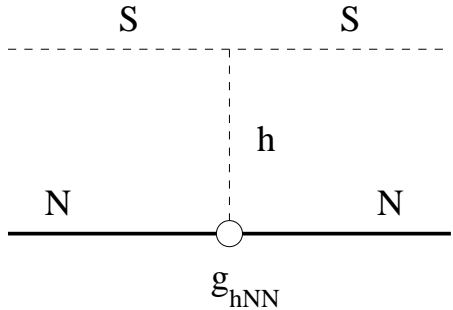}
%{\normalsize \mbox{\epsfxsize=60mm\epsffile{sNscatt.eps}} }
\end{center}

\caption{The Feynman graph relevant to scalar-nucleon elastic scattering.}
\end{figure}
%%%%%%%%%%%%%%%%%%%%%%%

The Feynman diagram describing elastic $S$-particle collisions with
nucleons and nuclei is given by $t$-channel Higgs exchange, as shown
in fig.~(3). If we write, for slowly-moving spin-$J$ nuclei, the relevant nuclear
matrix element as
\beq
\label{matrixelementdef}
\frac{1}{2J+1} \sum_{\rm spins}
\Bigl|\langle N'| \sum_f y_f \ol{f} f | N \rangle\Bigr|^2 
\approx \frac{|\ca_{\sss N}|^2}{(2 \pi)^6},
\eeq
then the nonrelativistic elastic scattering cross section obtained by
evaluating fig.~(3) becomes
\beq
\label{elasticsigma1}
\sigma_{\rm el} = \frac{\l^2 \vew^2 |\ca_{\sss N}|^2}{\pi}
\; \left( \frac{m_*^2}{\ms^2 \mh^4} \right) ,
\eeq
where $m_* = \ms \mn/(\ms+\mn)$ is the reduced mass for the collision.

Modelling the nucleus as $A$ independent nonrelativistic nucleons leads
to the expectation that $\ca_{\rm nucleus} \approx A \ca_{\rm nucleon}$, 
making it convenient to relate the elastic cross sections for nuclei and
nucleons by
\begin{equation}
\sigma_{\rm el}(\hbox{nucleus}) = \frac{A^2 m_*^2(A,S)}{m_*^2(p,S)} 
\; \sigma_{\rm el}(\hbox{nucleon}) , 
\label{sigmaE}
\end{equation}
The Higgs charge of the nucleon can be related to the nucleon mass and 
the trace anomaly, following ref.~\cite{SVZ}. The result is
sensitive to the mass of the strange quark and its content in the nucleon in the
0$^{+}$ channel. Taking the strange quark mass to be 170 MeV and 
$\langle N|\bar{s}s|N\rangle \simeq 0.7$, (see, for example, Ref. \cite{Z}), 
we deduce the estimate:
\begin{equation}
\ca_{\rm nucleon} = g_{\sss hNN}\approx \frac{340~{\rm MeV}}{\vew},
\end{equation}
leading to
\begin{eqnarray}
\sigma_{\rm el}(\hbox{nucleon}) &\approx& \left( \frac{\l ~340~{\rm MeV}}{\mh^2}
\right)^2~ \left( \frac{\mp}{\pi (\mp + \ms)} \right)^2 \nonumber\\
&=& \l^2 \left( \frac{100~{\rm GeV}}{\mh} \right)^4 
\left( \frac{50~{\rm GeV}}{\ms} \right)^2 ~ \Bigl( 20 \times 10^{-42}
~{\rm cm}^2 \Bigr)  .
\label{sisN}
\end{eqnarray}

\subsection{Recoil Experiments}

It is instructive to see what kind of cross section is implied by the
primordial cosmic abundance relation, $\l(\ms)$. Using $\mh = 120$ GeV
and $\ms = 40$ GeV, the abundance condition $0.1<\Omega_sh^2<0.3$
implies $ 0.05\la\l\la 0.085 $.
Using these in eqs.~\pref{sisN} then gives $ 5 \times 10^{-44}$ cm${}^2 \la
\sigma_{\rm el}(\hbox{nucleon})
\simeq 1.5 \times 10^{-43}$ cm${}^2$, which is about a factor of 20
smaller than the range of nucleon scattering cross sections 
which are accessible in the DAMA and CDMS
experiments \cite{CDMS,Dama}.

%%%%%%%%%%%%%%%%%%%%%%%
%%%%%%%%%%%%%%%%%%%%%%%
\begin{figure}[!t]
\begin{center}
\includegraphics[width=13.0cm]{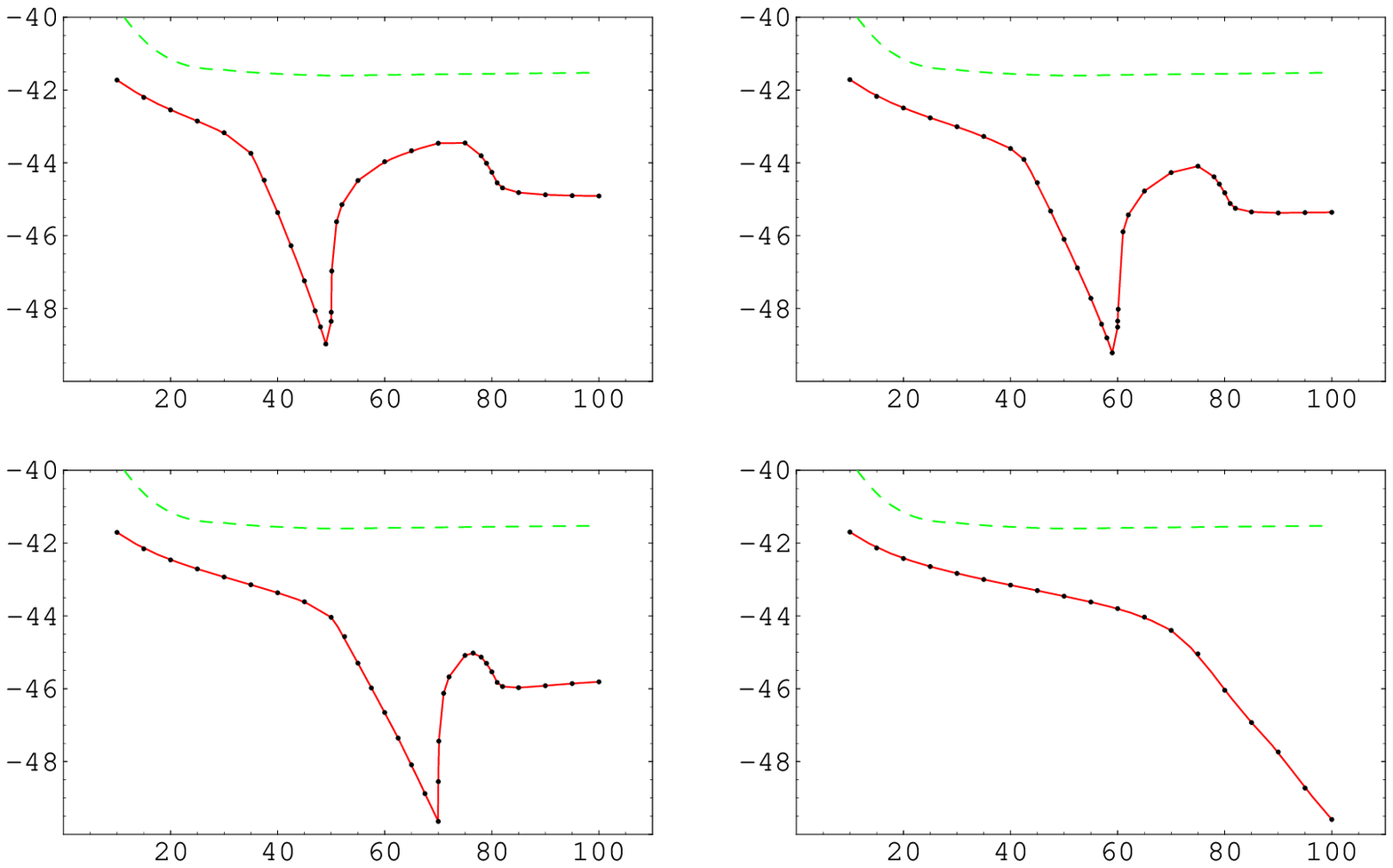}
\put(-197,80){\scriptsize{$m_S \, \left[ {\rm GeV} \right]$}}
\put(-197,0){\scriptsize{$m_S \, \left[ {\rm GeV} \right]$}}
\put(-67,80){\scriptsize{$m_S \, \left[ {\rm GeV} \right]$}}
\put(-67,0){\scriptsize{$m_S \, \left[ {\rm GeV} \right]$}}
\put(-237,97){\scriptsize{$m_h=100\, {\rm GeV}$}}
\put(-237,17){\scriptsize{$m_h=140\, {\rm GeV}$}}
\put(-107,97){\scriptsize{$m_h=120\, {\rm GeV}$}}
\put(-107,17){\scriptsize{$m_h=200\, {\rm GeV}$}}
\put(-258,100){\scriptsize{\rotateleft{$\log \sigma_{\rm el}({\rm nucleon}) 
  \left[{\rm cm}^2\right]$}}}
\put(-258,20){\scriptsize{\rotateleft{$\log \sigma_{\rm el}({\rm nucleon}) 
  \left[{\rm cm}^2\right]$}}}
\put(-129,20){\scriptsize{\rotateleft{$\log \sigma_{\rm el}({\rm nucleon}) 
  \left[{\rm cm}^2\right]$}}}
\put(-129,100){\scriptsize{\rotateleft{$\log \sigma_{\rm el}({\rm nucleon}) 
  \left[{\rm cm}^2\right]$}}}
\caption{The predictions for the elastic cross section, $\sigma_{\rm el}$, as 
a function of $\ms$, which follows from the $\l(\ms)$ dependence
dictated by the cosmic abundance. Also shown by a dashed line
is the exclusion limit from the CDMS experiment 
\protect\cite{CDMS} .}
%
%\psgrid(0,0)(-14,9)(1,-1)
\label{fig4}
\end{center}
\end{figure}

Fig.~(\ref{fig4}) shows how the scattering cross section we obtain in this
fashion is related to current experimental limits, under the standard
assumption that the mass density of $S$-particles in our galactic
halo is 0.3 GeV/cm${}^3$, having velocities of order 200 km/sec.
This plot shows the lowest possible values of the cross sections, 
as it uses the maximum possible abundance $\Omega_sh^2 =0.3$. For smaller 
values of the cosmic abundance, $\lambda^2$ is higher and 
the cross sections could be up to three times larger than those shown in 
fig.~(\ref{fig4}).
The predictiveness of the $\l-\ms$ relation makes this model much
easier to falsify than are more complicated models, with much of the
parameter space covered by the next generation of experiments \cite{Cad}.
Most importantly, the projected sensitivities of the CDMS-Soudan and Genius 
experiments will completely cover the range $\ms \leq 50$ GeV, for
values of the Higgs mass between 110 and 140 GeV. As we show in the 
next section, this range
of masses and coupling constants has important 
implications for the Higgs searches at colliders. On the other hand, there 
exists the possibility of completely ``hiding'' the dark matter by choosing
$0.4\mh\la \ms \leq 0.5\mh$. In this case annihilation at freeze-out
is very efficient, requiring small $\l$'s which lead to
elastic cross sections suppressed to the level of $10^{-48}$ cm$^2$. These 
levels of sensitivity to $\sigma_{\rm el}(\hbox{nucleon})$ are not 
likely to be achieved in the foreseeable future.

Our model of a singlet real scalar predicts a smaller signal for underground
detectors than does a model where the dark matter consists of
$N$ singlet scalars (including the model considered in ref. \cite{McD}, for 
which $N=2$). This is because the abundance of every individual species must 
be $1/N$ of the total dark matter abundance, $\Omega_i = \Omega_{\rm tot}/N$. 
This requires a larger annihilation rate at freeze-out for 
every species, and so an {\em enhancement} of the coupling $\lambda$ by 
$\sqrt{N}$, relative to the values of coupling constant calculated  
in this paper. By contrast, since the local halo density does not depend 
on $N$, the signal from an $N$-component scalar dark matter model
will be $N$ times larger than what is found here for
the signal from our single-component dark matter model. 

\subsection{High-Energy Neutrino Searches}

The elastic cross section calculated above, $\sigma_{\rm el}$, also
controls the expected flux of high-energy neutrinos which would be 
emitted by $S$ particles which are captured at the centre of the
sun or the earth. It does so because of the following scenario, which
describes how the abundance of such particles is determined. 

$S$ particles which lose enough energy through scattering get trapped in the
gravitational field of the sun or the earth. Further scattering permits them
to further dissipate their energy, until they eventually accumulate at the 
solar (or terrestrial) center. Their density at the centre grows until the 
accumulation rate precisely balances their rate of 
removal due to pair annihilation, 
leading to an equilibrium abundance. Because of this balance the total
rate of annihilations may be computed given the capture rate, and so also
given the elastic scattering rate. 

The detection of these captured particles is based on observing the high-energy
neutrinos which are among the production products of these annihilations. 
These neutrinos can escape from the solar (or terrestrial) centre, and
can be detected by neutrino telescopes, which look for the energetic muons 
which are produced when the neutrinos interact with rock or ice in the
detectors' immediate vicinity. In this way a signal is predicted for
detectors like Baksan, Kamiokande, Macro, Amanda 
and others.

To predict this signal we use the results of Ref. \cite{FOP}, which 
follows the original treatment in Refs. \cite{sos}, and estimate the 
neutrino flux at the surface of the earth due to $S$-particle annihilation in 
the sun as 
\begin{equation}
\phi _{\nu \odot }\simeq (560 \hbox{cm}^{-2} \hbox{s}^{-1}) \;
N_{\rm eff} \; \sigma_{\rm el,36} \; 
\frac{\mbox{GeV}^2}{\ms^2}.  
\label{flux}
\end{equation}
In this formula $N_{\rm eff}$ is the average number of 
neutrinos produced per
annihilation event, and 
$\sigma_{\rm el,36}$ is the $S$-proton elastic scattering
cross section in units of $10^{-36}$ cm$^{2}$. 

The determination of $N_{\rm eff}$ requires a study of the final states
which are available in $S$ annihilations, which are mediated by 
$s$-channel Higgs-exchange, as in Fig. (1). The number of neutrinos produced
depends on the value of $\ms$, which controls whether the main annihilation
products are $W^{+}W^{-}$, $b\bar{b}$ or other light hadrons. (Direct
branching to a pair of neutrinos is obviously very small for Higgs-mediated
decays.) The total production of energetic neutrinos turns out to be 
quite significant, as all decay products typically produce final state
neutrinos (and so non-vanishing $N_{\rm eff}$) due to cascades of
weak decays. The resulting value for $N_{\rm eff}$ and the energy spectra 
of the produced neutrinos have been meticulously simulated by Ritz and Seckel 
in Ref.~\cite{RS} and more recently by Edsjo \cite{Edsjo}. 

To obtain an upward-going flux of energetic muons, one must compute the
probability that a neutrino directed towards the detector produces a muon at
the detector. This probability, $P(E)$, is a quadratically 
rising function of neutrino energy, $E$, which must be convoluted with the computed
neutrino fluxes \cite{GS1,HM}. Approximating this probability 
as $P(E)\sim 10^{-8} (E/{\rm 100~GeV})^2$, we use the results of Ref. 
\cite{RS} for $60 $ GeV $b$ quarks injected into the center of the sun, which 
predicts $N_{\rm eff}$ to be $0.3$. The second moment of the neutrino
energy distribution, $N_{\rm eff}\, \langle (E/\ms)^2 \rangle$
becomes 0.006. Extrapolating these results to the case 
$\ms=50$ GeV gives
\beq
\phi_\mu = 0.8 \times 10^{-14} \left(
\frac{\sigma_{\rm el}(\hbox{nucleon})}{10^{-44}~{\rm cm}^2}\right) 
{\rm cm}^{-2}{\rm s}^{-1} .
\eeq
This should be compared with experimental limits on the flux of 
energetic upward muons (as obtained, for example, by the Kamiokande, 
MACRO and Baksan experiments, \cite{Kam,Mac,Bak}), which 
is $\phi_\mu \mathrel{\raise.3ex\hbox{$<$\kern-.75em\lower1ex\hbox{$\sim$}}}
1.4\times 10^{-14}$ cm$^{-2}$s$^{-1}$. Using the value of
$\sigma_{\rm el}(\hbox{nucleon})$
suggested by abundance calculations ({\it c.f.} eq.~\pref{sisN}), 
we can see that the predicted flux is
right at the level of current experimental sensitivity.  The 
constraints on the flux coming from the center of the earth can be 
equally important \cite{McD}.

There are, of course, caveats to the blanket use of these constraints, since
loopholes can exist for some values of the parameters. For instance, 
according to Ref. 
\cite{RS}, the average energy of the neutrinos is about 10\% of 
$\ms$, and so for lower values of $\ms$ the muons passing through the
detectors might be close to or below the experimental thresholds. We conclude 
that although the prospects are good for observable signals in these detectors,
more detector--oriented studies need to be done in order 
to exploit fully the limits which may result. 

\section{Implications for Collider Experiments}

We next turn to the implications of the model for Higgs searches at
colliders. As we saw from the primordial abundance constraints,
over most of parameter space $\l \simeq O(0.1)-O(1)$. This means that real or
virtual Higgs production might frequently also be associated with 
$S$ production, 
potentially leading to strong missing energy signals. (See 
refs.~\cite{invisible}
for discussions of the implications of related scalar models for 
collider experiments and refs.~\cite{missinghiggsatsusy}
for the discussion of invisible Higgs decay in supersymmetric models.) 
Since the extent to which this actually happens 
depends strongly on the $S$-particle mass, we consider the main alternatives 
successively.

\subsection{$2m_s<m_h$} 

This is the most interesting possibility, since it permits the
decay of the Higgs into a pair of $S$-particles. The decay width,
calculated at tree level, is given by 
\begin{equation}
\Gamma_{h\rightarrow SS} = 
\frac{\l^2 \vew^2}{8\pi \mh} \sqrt{1-\frac{4\ms^2}{\mh^2}}
\qquad {\rm for} \qquad 2 \ms < \mh.
\end{equation}
Needless to say, since any produced $S$ particles are unlikely to interact
within the detectors, the large values for $\l$ suggested by
cosmic abundance can give a very large invisible decay width to the Higgs,
without changing the Higgs' other Standard Model couplings.
This fact obviously has many consequences for Higgs searches at 
colliders, since Higgses will be produced at the rates expected in the
Standard Model, but will mostly decay invisibly into $S$ pairs. 
To quantify this observation we calculate the ratio of the above 
invisible width to the width for the decay $h \to \tau^+\tau^-$:
\begin{equation}
\frac{\Gamma_{\sss h\rightarrow SS}}{\Gamma_{\sss h\rightarrow \tau^+\tau^-}}=
\frac{\l^2 \vew^4}{\mh^2 m_\tau^2} \sqrt{1-\frac{4\ms^2}{\mh^2}}.
\end{equation}
Using the usual cosmic abundance constraints, obtained in Section 3, one 
can see that this ratio can be as large as 1000, so that the decay
of the Higgs 
into a pair of $S$ scalars is by far the more probable event.  
More useful information is given, however, by another ratio,
\be
R=\frac{{\rm Br}_{h\rightarrow \tau^+\tau^-}(SM+S)}
{{\rm Br}_{h\rightarrow \tau^+\tau^-}(SM)}
=\frac{\Gamma_{h,~total}(SM)}{\Gamma_{h\rightarrow SS}+\Gamma_{h,~total}(SM)} ,
\label{downgrade}
\ee
where the last equality uses the fact that Higgs production 
rates are not affected 
by the existence of a new scalar. $R$ quantifies the 
deterioration (relative to the
SM result) of the expected signal for Higgs decaying into 
visible modes due to the 
adding of the new scalar. 
We plot $R$, in fig.~(5), against $\ms$ up to the Higgs mass for the same 
values of $\mh$ and $\lambda$ as in fig.~(2). 

%%%%%%%%%%%%%%%%%%%%%%%
%%%%%%%%%%%%%%%%%%%%%%%
\begin{figure}[!t]
\begin{center}
\includegraphics[width=13.0cm]{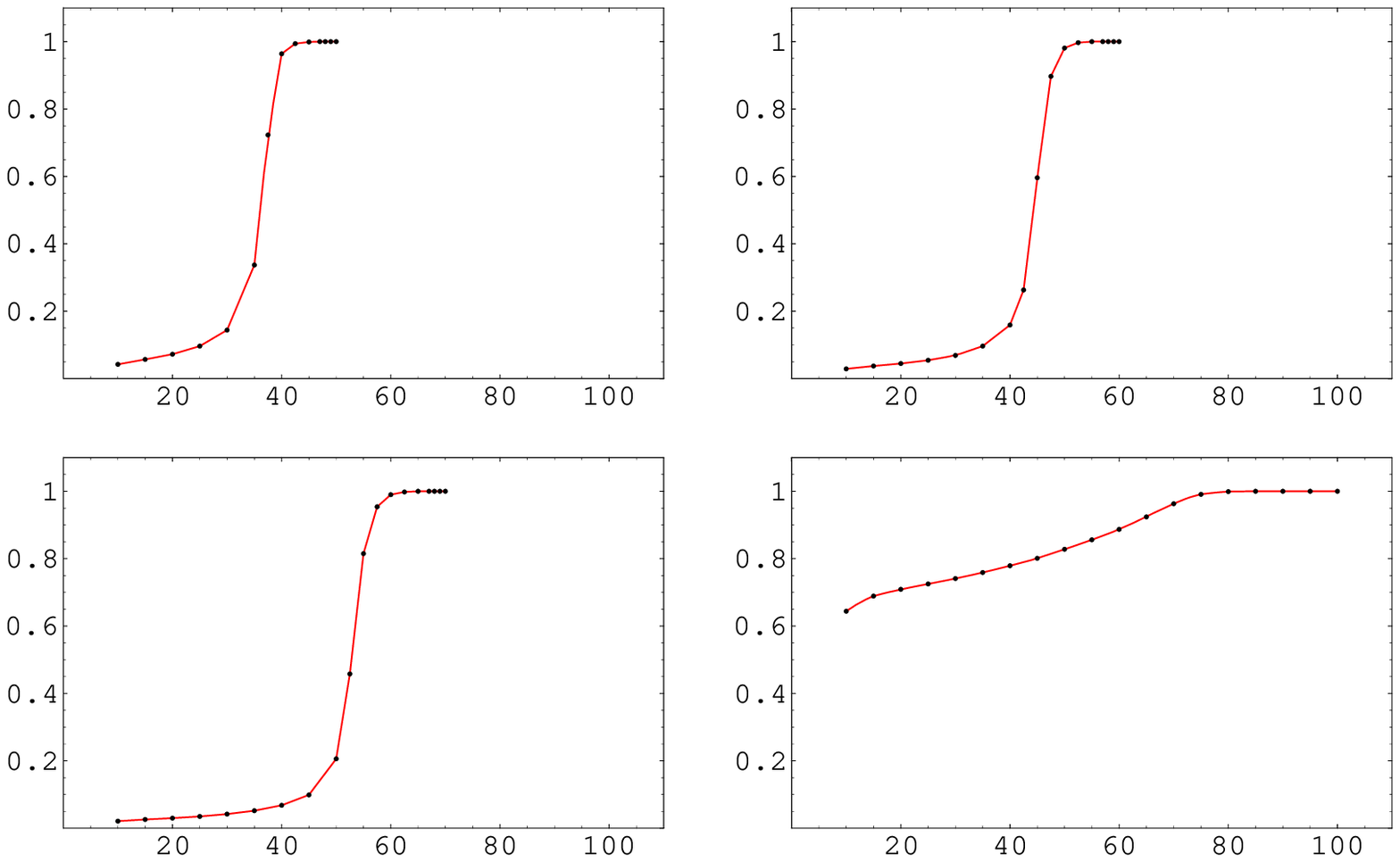}
\put(-197,80){\scriptsize{$m_S \, \left[ {\rm GeV} \right]$}}
\put(-197,0){\scriptsize{$m_S \, \left[ {\rm GeV} \right]$}}
\put(-67,80){\scriptsize{$m_S \, \left[ {\rm GeV} \right]$}}
\put(-67,0){\scriptsize{$m_S \, \left[ {\rm GeV} \right]$}}
\put(-180,97){\scriptsize{$m_h=100\, {\rm GeV}$}}
\put(-180,17){\scriptsize{$m_h=140\, {\rm GeV}$}}
\put(-50,97){\scriptsize{$m_h=120\, {\rm GeV}$}}
\put(-50,17){\scriptsize{$m_h=200\, {\rm GeV}$}}
\put(-258,119){\scriptsize{\rotateleft{${\rm R}$}}}
\put(-258,39){\scriptsize{\rotateleft{${\rm R}$}}}
\put(-129,39){\scriptsize{\rotateleft{${\rm R}$}}}
\put(-129,119){\scriptsize{\rotateleft{${\rm R}$}}}
\caption{The ratio, $R$, of the total Higgs width 
in the Standard model over the same width in the Standard Model supplemented
by the singlet scalar, plotted as a function of $\ms$. }
%
%\psgrid(0,0)(-14,9)(1,-1)
\label{fig5}
\end{center}
\end{figure}

As is clear from the plot, for $m_h = 100$, 120 and 140 GeV and $2\ms < \mh$
the invisible width dominates the total 
width everywhere except in the immediate vicinity of $2\ms=\mh$. $R$ shrinks
near this point for two reasons. First, $\mh$ near $2\ms$ is close to
threshold for producing two $S$ particles in $h$ decay, and so the invisible 
rate is phase-space suppressed in this region. More importantly, the size
of the coupling, $\l$, allowed by abundance arguments is smaller for $\ms$
in this region, due to the enhancement of the primordial annihilation cross section
due to proximity with the Higgs pole. 

However, the plot also shows that the presence of the invisible width already
downgrades the expected signal by a factor of 10 or more when $m_S \simeq 0.3 \mh$ GeV 
for $\mh = 120$ or 140 GeV. This means that over 10 times more Higgs particles 
must be produced in order to reach the same level of signal 
in all visible decay modes, as compared to what is expected purely within
the Standard Model. This implies a 
tremendous suppression of the observable Higgs
signal at the LHC and especially at the Tevatron --- 
possibly even precluding its 
discovery at these machines (see, however, ref.~\cite{Gunion,MW,Dieter} 
for a more
optimistic point of view). 
The main Higgs production reaction in this case
would be two jets plus missing energy, a process having enormous backgrounds
in hadron machines. 

On the other hand, for $\mh$ close to or above the $W^+W^-$
threshold, the similarity in size between 
$\l$ and the electromagnetic coupling,
$e$, implies that the invisible width does not completely dominate other decay 
modes. 
In such a case (fig. (5), $\mh=$200 GeV) the decline in $R$ 
is not dramatic, and the existence of an invisible signal
need not preclude finding the Higgs in a hadron machine.

More can be said about an invisibly decaying Higgs at $e^+e^-$ machines.
Although the missing energy signal would be missed in an orthodox Standard-Model
Higgs search, such as one based on $b$-tagging, more model-independent
searches have been performed for scalars produced with Standard Model 
cross sections, but decaying invisibly. These searches at LEP exclude such an
invisibly-decaying Higgs unless $\mh \ga 106.7$ GeV \cite{LEPBounds}. 

\subsection{$2 \ms > \mh$}

For this mass pattern, $S$ particles cannot be produced by real 
Higgs decays, and
so arise only through virtual Higgs exchange. Again, once produced,
these particles are not expected to interact within the detector and so look
like missing energy above an energy threshold, $E\geq 2 \ms$.

A missing-energy signal can be searched for at $e^+e^-$ colliders, where the
backgrounds are very well understood. Since hadronic machines are unlikely
places for seeing such a signal, we do not consider them further here. 

%%%%%%%%%%%%%%%%%%%%%%%
\begin{figure}[hbtp]
\begin{center}
\includegraphics[width=9.0cm]{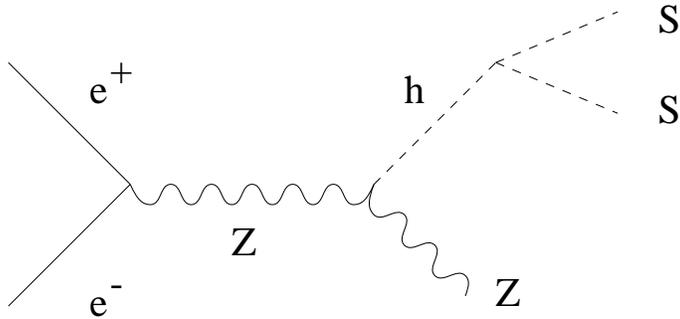}
%{\normalsize \mbox{\epsfxsize=90mm\epsffile{missingEt.eps}} }
\end{center}

\caption{The dominant mechanism for pair-producing $S$-particles at
LEP, via $s$-channel Higgs exchange.}
\end{figure}
%%%%%%%%%%%%%%%%%%%%%%%

Fig.~(6) shows the main mechanism for the production of two $S$ scalars 
at LEP (or at a linear collider). The differential cross-section for this 
process which results from the evaluation of this graph is as follows:
\begin{eqnarray}
{\frac{d\sigma_{\sss e^+e^-\rightarrow ZSS}}{dt~du}} = 
\frac{(g_{\sss L}^2+g_{\sss R}^2) \l^2 \vew^2 |a_z|^2}{16 (2\pi)^3 s} 
~ \left(\frac{e}{\sw\cw}\right)^2  \left[ {\frac{s+ (t-\mz^2) 
(u-\mz^2)/\mz^2 }{(s- \mz^2)^2+ \Gamma_{\sss Z}^2 \mz^2}} \right]  \nonumber \\
\times \left[{\frac{1}{s+t+u-\mz^2-\mh^2}} \right ]^2
\sqrt{1-\frac{4 \ms^2} {(s+t+u- \mz^2)}}. 
\label{missingLEP}
\end{eqnarray}
Here we follow the notation of ref.~\cite{BMP}, with $g_{\sss L} = -\, \frac12+
\sw^2$ and $g_{\sss R} = \sw^2$ parameterizing charged-lepton couplings 
to the $Z$ boson, 
$a_z = e \mz /(\cw \sw)$ being the effective coupling of the Higgs to $Z$ boson, 
and $\sw$ and $\cw$ denoting the sine and cosine of the Weinberg angle. 
The Mandelstam variables, $s$, $t$ and $u$, are defined as for a
2-body to 2-body process, with their sum given by $s+t+u = \mz^2 + Q^2$,
where $\mz$ is the $Z$-boson mass and $Q^2$ is the square of the invariant mass
of the two $S$-particles.

Expression \pref{missingLEP} also applies to the previous case, where $2\ms < \mh$, in
which case the cross section is dominated by the Higgs pole, corresponding
to real Higgs production. It is clear that the constraint on the missing energy 
implied by eq.~(\ref{missingLEP}), and the experimentally measured values for such 
a process at LEP, cannot lead to a limit on the invariant mass of two $S$ 
particles better than 100 GeV for a Higgs mass of around 100 GeV. 
For larger values of $\mh$ a possible bound on the invariant
mass of the $S$ pair is considerably relaxed. 

\section{Conclusions}

We have presented the first study of the minimal model for non-baryonic
dark matter, which consists of the Standard model plus a singlet scalar. 
This model is characterized by one additional real scalar field and 
three new parameters. 
The absence of linear and cubic terms in $S$ is required to ensure that the
the new scalar is sufficiently stable to contribute significantly to
the dark matter currently present in the Universe. This stability also
precludes the development
of a nonzero v.e.v. for $S$. The study of the model's scalar potential 
shows that electroweak symmetry is spontaneously broken while $\langle S
\rangle$ vanishes for a significant domain of parameter space.

At the renormalizable level the three parameters describe the $S$ particle
mass, its self-coupling and its couplings to other Standard Model fields,
which all are mediated by a coupling to the Standard Model Higgs,
of the form $\l S^{2}(H^{\dagger }H)$. The simplicity of this
coupling significantly simplifies the calculations of primordial abundances and
observable signals at colliders and underground detectors, making the resulting
predictions more certain. The primordial abundance is governed mainly by the
annihilation of $S$-particles via $s$-channel Higgs exchange. 
If the mass of the Higgs is not too close to $2\ms$, the
observed abundance of dark matter is achieved by a most natural choice for
the coupling, $\lambda \sim O(0.1 - 1)$. 

These large values of $\lambda$ lead to nuclear elastic cross sections 
which are $ \sim 10^{-43}$ cm$^2$  in size. This is slightly below the limits
of sensitivity of the DAMA and CDMS experiments, but is likely to be detectable
at future experimental facilities. Thus, the direct detection of $S$
particles is not yet ruled out, and is easily feasible within the near 
future! Our minimal model 
is unable to reconcile the DAMA and CDMS results, and so predicts that 
one or the
other is incorrect. 

Cross sections this large also lead to a
significant flux of high-energy neutrinos generated by $S$ annihilation
at the centers of the earth and the sun, at levels potentially accessible to
large neutrino detectors. We give here only an estimate of the expected flux, 
and we believe the encouraging results motivate further
work to simulate the expected neutrino and muon spectra and
intensities.
It would also be intriguing to repeat the numerical analysis
of ref. \cite{McD}, updated with the Higgs masses favored by modern collider results.

Large values for $\l$ may also lead to the significant missing-energy signals
at colliders, corresponding to the production of a pair of $S$-particles. 
This signal is unlikely to be seen at hadronic machines, where
the background events (two jets plus missing energy) will have much larger
cross-sections. Lepton colliders, such as LEP or NLC are needed. The possible
existence of $S$-particles with a mass smaller than half of the Higgs
boson mass poses a significant threat to Higgs searches at the Tevatron and LHC.
Indeed, in this case the invisible decay of the Higgs boson can be many
orders of magnitude larger than the search modes, 
making the Higgs effectively invisible. For 
intermediate Higgs masses, up to 140 GeV and higher, the pair
production of $W$ bosons introduces a nonnegligible visible width
and so saves the Higgs searches. It is important that in the
dangerous domain of parameter space, $2 \ms < \mh < 140$ GeV, the
size of the elastic cross-section, $\sigma_{\rm el}$, is limited {\em from below}
to be larger than $10^{-44}$ cm$^2$. If this were indeed the case, the next
generation of the underground detectors would likely discover the recoil
signal.

Because the self-coupling, $\ls$, is unconstrained by the abundance
condition, one might hope to use models of this sort to produce a
strongly-interacting dark matter candidate, a possibility recently advocated 
in ref.~\cite{selfint,BBRT}. Unfortunately, this
proposal would also require a rather small mass for the $S$-scalars (1 GeV or less),
and since $\l$ cannot be small for such masses, one is led to a fine-tuned (ppm) 
cancellation between $\lambda v_{EW}^{2}$ and the ``bare'' mass $m_{0}^{2}$. 
Worse, having $\ms$ be 500 MeV or smaller pushes $\l$ to the very limits
of our perturbative 
analysis. 

It remains an interesting challenge to construct a non-minimal 
model of strongly-interacting dark matter along the present lines. 
One possibility is to suppress $\l$ to evade the fine-tuning problem, and 
set the freeze-out abundance by a set of non-renormalizable operators of
dimension 6 and higher.  Another possibility for
having light and strongly interacting scalar dark matter  
requires $\l$ to be very small, scalar particles to be out of equilibrium 
at all temperatures, and their abundance to be fine-tuned 
(for example,  using inflationary scenarios). The suppression of the 
coupling to the Higgs, strong self-interaction and a low mass for $S$ 
could be simultaneously achieved in models where $S$ is 
composite \cite{FP}, in which case there is unlikely to be 
significant impact on Higgs physics. 

Our pursual here of a more model-independent approach to the 
physics of dark matter particles is ultimately stimulated by the maturity of 
current dark-matter detection efforts. As the data erodes the large parameter 
space of the popular neutralino models, or if a dark matter candidate is
eventually found, a reliable interpretation of the experiments can only be 
found by comparing with a wide variety of explanatory models. Our proposal in
this paper marks the absolute minimum of complexity required by
such a model, and from this minimality springs its predictiveness. It is also
generic in the sense that it is the most general renormalizable model consistent with the
assumed particle content and symmetries. As such, it captures the implications of any 
models whose implications for dark matter arise from a low energy spectrum which
contains only these particles and symmetries. 
We believe that further
model--building efforts are in order as experimental developments progress.

\bigskip

\bigskip

\bigskip

\centerline{\bf Acknowledgements}
\bigskip
We thank G. Couture, T. Falk, L. Kofman, 
G.Mahlon, K. Olive, S. Rudaz and M. Voloshin 
for usefull discussions and comments. M.P. wishes to thank the 
Physics Departments
of McGill University and the University of Qu\'ebec at Montr\'eal,
where a significant portion of this work was carried out. M.P. and
T. t V.'s research was funded by the US Department of Energy under
grant number DE-FG-02-94-ER-40823, and C.B. acknowledges the support
of NSERC (Canada), FCAR (Qu\'ebec) and the Ambrose Monell Foundation.

\newpage

\def\NPB#1#2#3{{\it Nucl. Phys.} {\bf B{#1}} (#2) #3}
\def\PLB#1#2#3{{\it Phys. Lett.} {\bf B{#1}} (#2) #3}
\def\PRD#1#2#3{{\it Phys. Rev.} {\bf D{#1}} (#2) #3}
\def\PRL#1#2#3{{\it Phys. Rev. Lett.} {\bf {#1}} (#2) #3}
\def\PRT#1#2#3{{\it Phys. Rep.} {\bf {#1} C} (#2) #3}
\def\MODA#1#2#3{{\it Mod. Phys. Lett.} {\bf {#1}} (#2) #3}

\end{document}